# Fiber-Optic Communication and Torsion Sensing via a BB84-Inspired Polarization Scheme

Vinícius Piaia, Rodrigo Cosme, Paulo Robalinho, Susana Silva, Henrique M. Salgado, and Orlando Frazão

*Abstract*— **Polarization-encoded optical-fiber links provide a useful platform for investigating the coexistence of communication and sensing functions in a shared optical channel. In this work, a classical analog of the BB84 polarization-encoding scheme was implemented over a bend-insensitive fiber link to evaluate how temperature, axial strain, and torsion affect the transmitted state of polarization and the recovery of a binary key. Temperature variations from 24 °C to 35 °C and axial strain up to 2 mε produced only minor polarization-state variations and did not affect key recovery. In contrast, fiber torsion induced a pronounced retarder-like evolution of the state of polarization, reducing the detector-intensity contrast used for bit discrimination and degrading key recovery. For intermediate torsion angles, the erasure rate reached 50%. The polarization state tended to recover for rotations close to Δθ = 180°, consistent with the oscillatory response observed on the Poincaré sphere. These results identify torsion as the dominant impairment in the tested polarization-encoded fiber link and demonstrate its potential use as a sensing mechanism in integrated communication-and-sensing architectures.**

## I. INTRODUCTION

THE development of optical-fiber telecommunications has enabled high-capacity transmission channels capable of supporting hundreds of gigabits per second over a single fiber. Beyond data transmission, optical fibers also provide intrinsic sensing capabilities, since their propagation properties are affected by external parameters such as temperature, vibration, strain, and torsion.

Integrated sensing and communication architectures are therefore under active investigation [1], particularly in scenarios where the same optical link may be used both to transmit information and to detect environmental or mechanical perturbations. Such concepts are also relevant to emerging quantum-network architectures [2]– [4], in which the stability of the optical channel is critical for preserving the encoded information. Quantum key distribution (QKD) is strategically important in this context because it enables key exchange based on physical principles rather than computational assumptions. Understanding how environmental perturbations affect polarization-encoded links is therefore important for the development of robust quantum-inspired and quantum-enabled fiber networks.

Unlike classical cryptographic schemes such as RSA, whose security depends on the computational difficulty of mathematical problems, the BB84 protocol proposed in 1984 [5] encodes binary information in nonorthogonal single-photon polarization states. In fiber-based implementations, however, polarization evolution due to birefringence and external perturbations can degrade the agreement between Alice and Bob. Long-distance QKD over optical fibers [6] has motivated several detection and implementation approaches, including transition-edge sensors [7], photon-interference B92/BB84 demonstrations [8], Yuen-type protocols [9], and subcarrier-wave BB84 systems [10].

In this work, a classical polarization-encoding analog inspired by the BB84 basis structure is implemented to evaluate how controlled physical perturbations affect polarization-state preservation and binary key recovery in an optical-fiber link. The proposed scheme is not intended for quantum key distribution, but rather as a classical test platform to assess the sensitivity of polarization-encoded transmission to channel perturbations relevant to future quantum-compatible fiber systems.

The main contribution of this work is the experimental evaluation of the effects of temperature, strain, and torsion on a classical BB84-inspired polarization-encoded fiber link. While temperature and strain had negligible impact within the tested ranges, torsion strongly altered the state of polarization, reduced the bit-discrimination contrast, and partially erased the recovered key. These results identify torsion as the dominant impairment and as a potential sensing mechanism for integrated communication-and-sensing fiber systems.

## II. EXPERIMENTAL ARRANGEMENT

Information transmission was carried out through a 2-m section of BendBrightXS bend-insensitive optical fiber from Draka, with an attenuation of 0.188 dB/km at 1550 nm. This macrobending-enhanced single-mode fiber (SMF), originally developed for fiber-to-the-home (FTTH) networks, offers improved bend tolerance, higher mechanical resistance to strain, and low birefringence, making it suitable for stable polarization-based transmission experiments. The experimental setup is illustrated in Fig. 1.

Manuscript received [date]; revised [date]; accepted [date]. This work was supported by No funding. (Corresponding author: Orlando Frazão.)

Vinícius Piaia, Rodrigo Cosme, Paulo Robalinho, Susana Silva, Henrique M. Salgado, and Orlando Frazão are with INESC TEC - Institute for Systems and Computer Engineering, Technology and Science, Porto, Portugal (e-mail: vinicius.piaia@inesctec.pt; orlando.frazao@inesctec.pt).

Vinícius Piaia is also with FEUP - Faculty of Engineering of the University of Porto, R. Dr. Roberto Frias, 4200-465, Porto, Portugal.

Rodrigo Cosme is also with FCUP - Faculty of Science of University of Porto, Rua do Campo Alegre, 4169-007, Porto, Portugal.

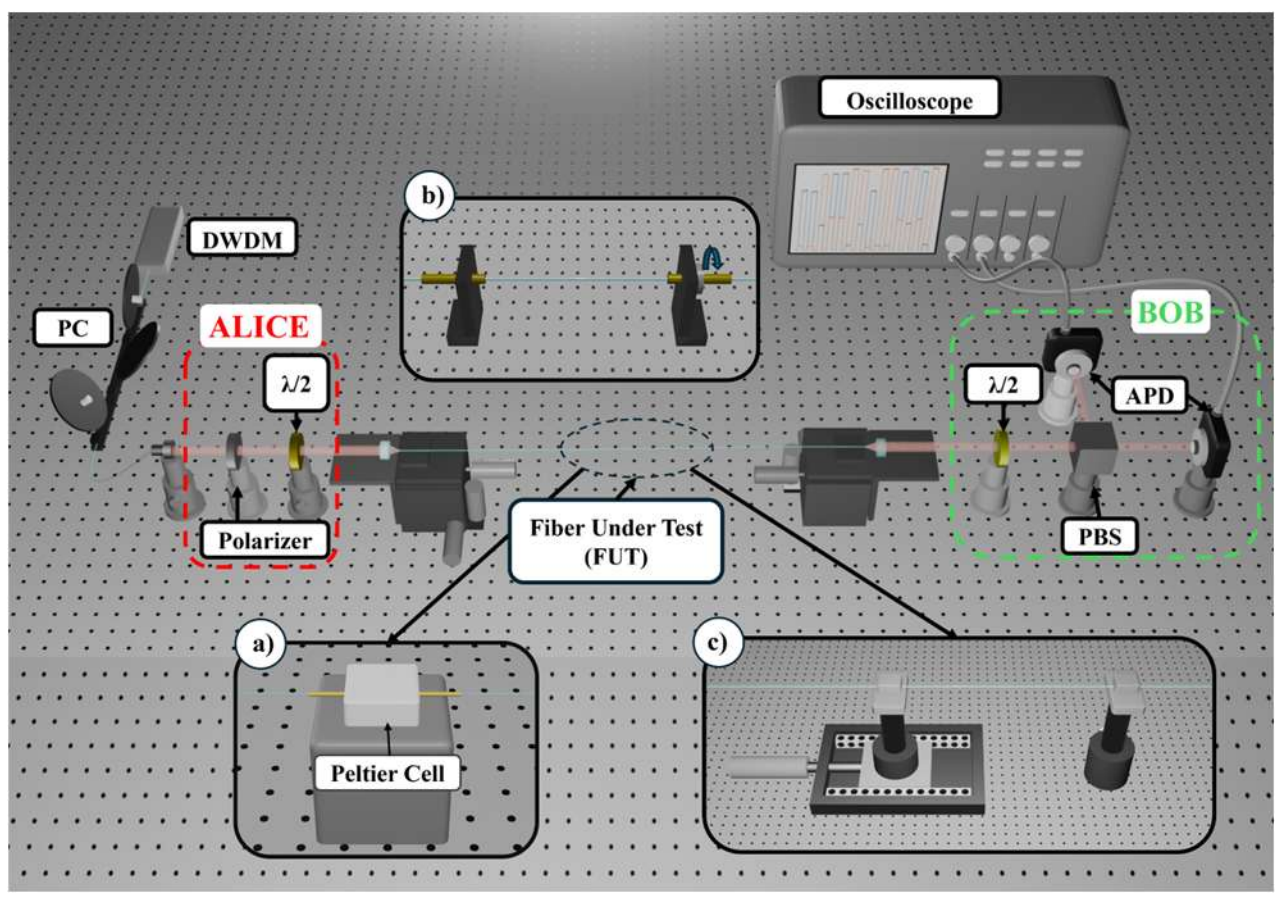

**Fig. 1.** Experimental setups employed to subject the FUT to different perturbations: (a) temperature variation, (b) torsional loading, and (c) axial strain.

A dense wavelength-division multiplexing (DWDM) laser source operating at 1550.12 nm (XYTSharetop) was used as the optical source. The emitted light was first guided through a standard single-mode fiber (SMF-28), passed through a polarization controller, and then collimated. A linear polarizer was placed before the encoding stage to improve the definition of the input polarization state.

The transmitter, referred to as Alice, was implemented using a half-wave plate. By adjusting its angular position, both the encoding basis and the transmitted bit value were selected. The rectilinear and diagonal polarization bases were generated through appropriate rotations of this wave plate. After the encoding stage, the optical signal was coupled into the BendBrightXS fiber under test using an 8 mm focal-length converging lens. Controlled perturbations were then applied to the fiber section to evaluate the robustness of the polarization-encoded communication channel and assess the feasibility of using the same fiber link for sensing.

At the end of the transmission link, the light was collected with another 8 mm converging lens and collimated toward Bob. The measurement basis was selected by adjusting the second half-wave plate. The light was then guided to a polarizing beamsplitter, where bit 0 (horizontal polarization) and bit 1 (vertical polarization) were separated. The two outputs were detected by avalanche photodetectors (APD431C, Thorlabs) connected to an oscilloscope (GDS-2304A, GW Instek).

## III. EXPERIMENTAL RESULTS AND ANALYSIS

A classical analysis of the BB84 protocol can be performed by treating the polarization state of light as the information carrier. When Bob receives the optical signal, it is directed to the photodetector associated with bit 0 or bit 1. The signal is monitored on an oscilloscope, and the bit value is assigned to the photodetector with the highest optical intensity. When Alice and Bob use different bases, Bob cannot unambiguously distinguish between bits. In the classical implementation considered here, this case is observed as similar intensity levels at the photodetectors associated with both bits. In an ideal implementation without perturbations or eavesdropping, full agreement between the sifted keys generated by Alice and Bob would be expected after basis reconciliation.

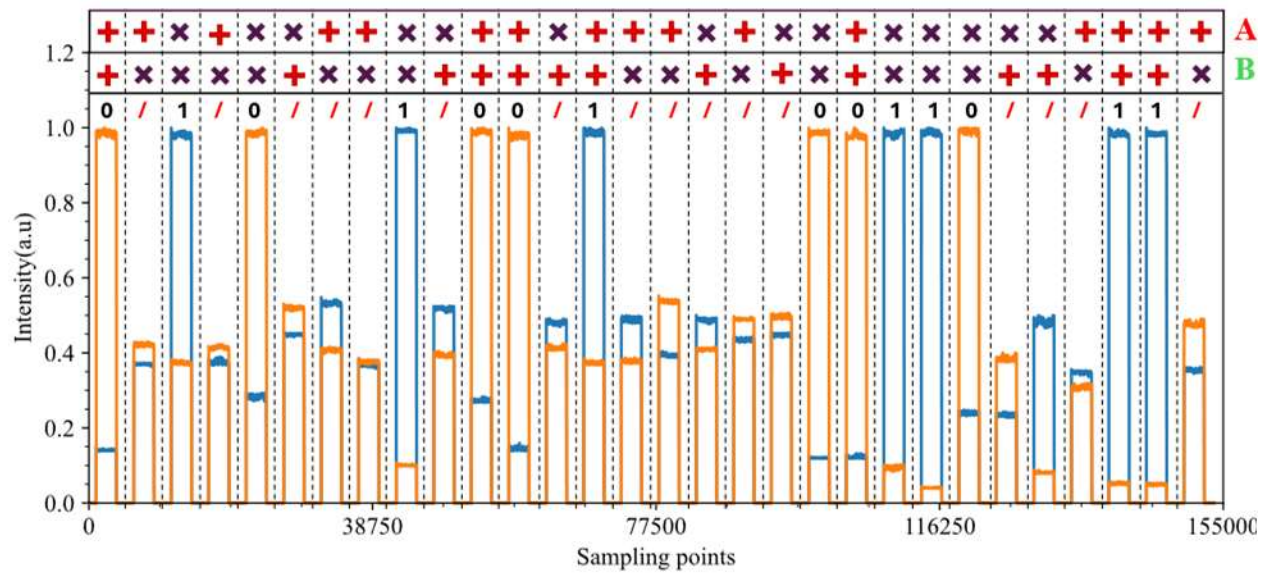

**Fig. 2.** Generated key in the absence of physical perturbations on the fiber under test.

In practice, physical perturbations along the communication channel may alter or degrade the transmitted polarization state, compromising the encoded information. Understanding these effects is therefore essential for reliable polarization-encoded communication. Fig. 2 shows key generation without physical perturbations, using a 30-bit raw key. The accepted bits were selected when Alice's and Bob's bases matched, yielding a 14-bit sifted key: 01010010011011. This key was used as a reference to evaluate the impact of controlled physical perturbations. Key-recovery performance was assessed by comparing Alice's and Bob's selected bases and bits. In parallel, the state of polarization after fiber propagation was characterized using a polarimeter (PM1000, Novoptel), providing insight into channel-induced polarization changes and their impact on key transmission.

To provide a quantitative comparison between the applied perturbations, the polarization-state variation ΔS (1) was quantified as the accumulated geodesic path length on the normalized Poincaré sphere. Considering consecutive normalized Stokes vectors $\mathbf{S}_i = (S_{1,i}, S_{2,i}, S_{3,i})$, the trajectory length was calculated as the sum of the angular distances between adjacent points, and erasure rate, defined as $ER=N_{erased}/N_{total}$, was also evaluated for each test condition, where $N_{erased}$ is the number of bits rejected due to failing the signal contrast threshold, and $N_{total}$ is the total number of transmitted bits. In addition to the erasure rate, the bit-error rate ($BER=N_{error}/N_{total}$) was also evaluated to distinguish between rejected bits and incorrectly recovered bits (where $N_{error}$ represents the number of mismatched bits between the transmitted and recovered sifted keys).

$$\Delta S = \sum_{i=1}^{N-1} \arccos(\boldsymbol{S}_i \cdot \boldsymbol{S}_{i+1}) \qquad (1)$$

### *A. Temperature and Strain-Induced Impact on the Communication Channel*

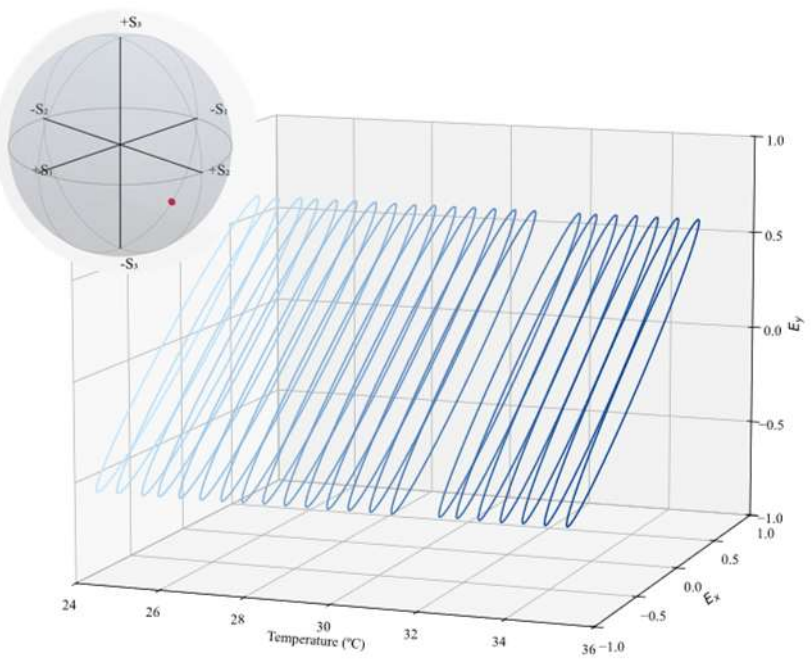

**Fig. 3.** Temperature effect on the polarization state.

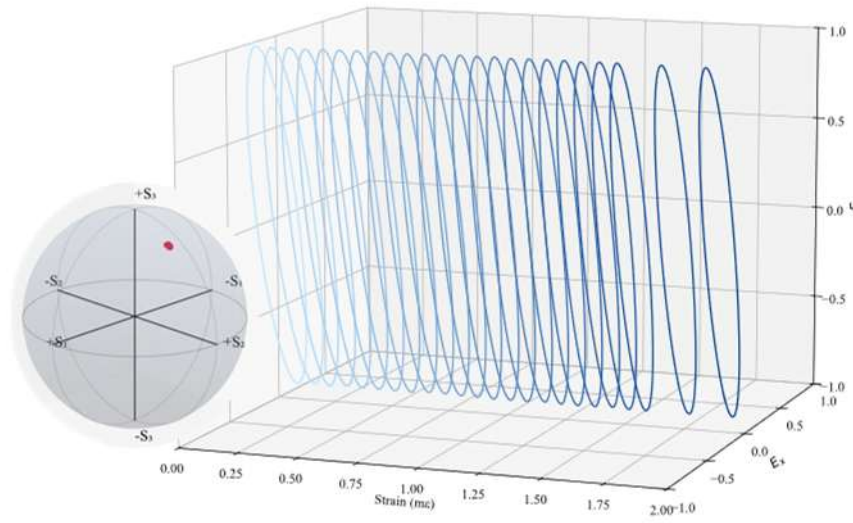

**Fig. 4.** Strain effect on the polarization state.

Within the tested ranges, temperature and strain did not introduce measurable noise into the detected signal or noticeably affect the communication channel (Table I). This result suggests that the polarization-encoded link may coexist with temperature- or strain-related monitoring functions, provided that the induced perturbations remain within comparable limits.

Temperature evaluation was conducted by placing the optical fiber in contact with a Peltier cell and varying the temperature from 24 *°C* to 35 *°C*. The polarization remained essentially unchanged over this interval, and key transmission was therefore unaffected, demonstrating insensitivity to the tested thermal variation, as shown in Fig. 3.

Strain was applied between two fixed points separated by 27.5 *cm* using a translation screw, up to a maximum strain of 2 *mε*. The polarimeter results, shown in Fig. 4, indicate low sensitivity of the polarization state to strain, which also did not influence key transmission. This stability may be associated with the low residual birefringence typically observed in weakly birefringent single-mode fibers, which limits differential changes in the optical paths associated with the $E_x$ and $E_y$ field components.

TABLE I

*ER* Evaluation due to Temperature and Strain

| Parameter | *Recovered Key* | *ER* |
|---|---|---|
| Temperature | 01010010011011 | 0% |
| Strain | 01010010011011 | 0% |

*B. Torsion-Induced Distortion of the Communication Channel*

Angular torsion was applied between two fixed points separated by 27.5 cm. When a single-mode fiber is twisted, the local birefringence axes rotate along the propagation direction, modifying the relative phase and orientation of the optical field [11]. In the present experiment, torsion produced a clear change in the measured state of polarization, exhibiting a retarder-like response similar to that of a quarter-wave plate, as shown in Fig. 5. This effect also influenced key transmission, because torsion reduced the intensity difference between the bit-0 and bit-1 detectors and, at critical points, fully compromised key recovery.

An oscillatory response was observed as the optical fiber was subjected to successive half-turn torsions. For a total rotation of approximately 180°, the state of polarization tended to return near its initial condition. This behavior is illustrated on the Poincaré sphere in Fig. 5, where the optical field evolves from a linear polarization state toward circular states and subsequently returns to a linear state. The difference in polarization variation caused by the physical parameters is illustrated in Table II.

TABLE II

*ΔS* Variation Path Trajectory on the Poincaré Sphere

| Parameter | *ΔS* |
|---|---|
| Strain (0 - 2 mε) | 0.39 ± 0.01 |
| Temperature (24 - 35ºC) | 0.29 ± 0.01 |
| Torsion Rate (0 – 11.4 rad/m) | 8.97 ± 0.01 |

## IV. KEY TRANSMISSION

Because the bits are encoded in the state of polarization, key recovery depends directly on the preservation of this state along the communication channel. Among the applied perturbations, only torsion produced a clear influence on the key. A dedicated evaluation was therefore performed to determine how the recovered key changes with the applied torsion angle.

A random sequence of bits was transmitted from Alice to Bob while different total torsion angles, ranging from 0° to 180°, were applied to the optical fiber. Considering the torsion rate defined as $\tau = \Delta\theta/L$ and a perturbed length of $L = 27.5$ *cm*, this interval corresponds to approximately 0 to 654.5°/*m*, or 0 to 11.4 *rad/m*. As shown in Fig. 6, the recovered key was affected by the applied torsion even in the absence of Eve. These results demonstrate that physical perturbations in the communication channel can compromise the transmitted information independently of an eavesdropping attack.

During key transmission, specific measurement bases were affected by the applied torsion. For Δθ = 45° and 135°, the bits associated with the perpendicular basis became indistinguishable, preventing reliable discrimination. In contrast, for Δθ = 90° and 135°, the diagonal basis was predominantly affected. This indicates that fiber torsion modifies the transmitted polarization state in a basis-dependent manner, degrading the agreement between Alice's and Bob's keys.

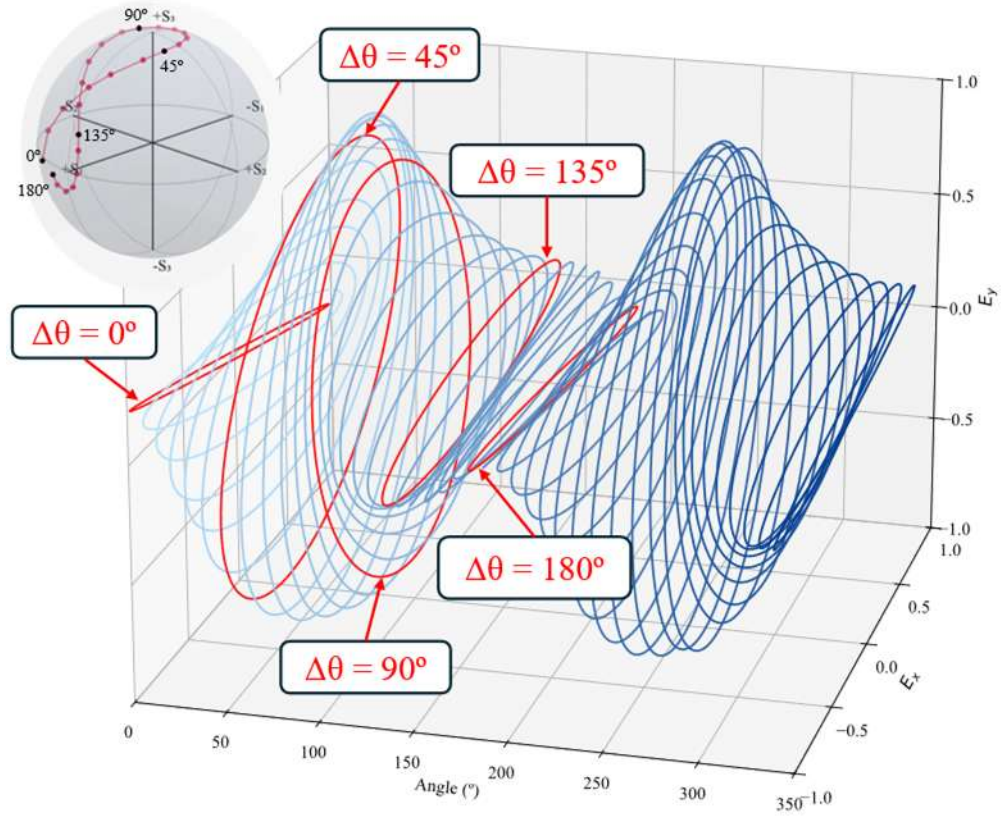

**Fig. 5.** Torsion effect on the polarization state.

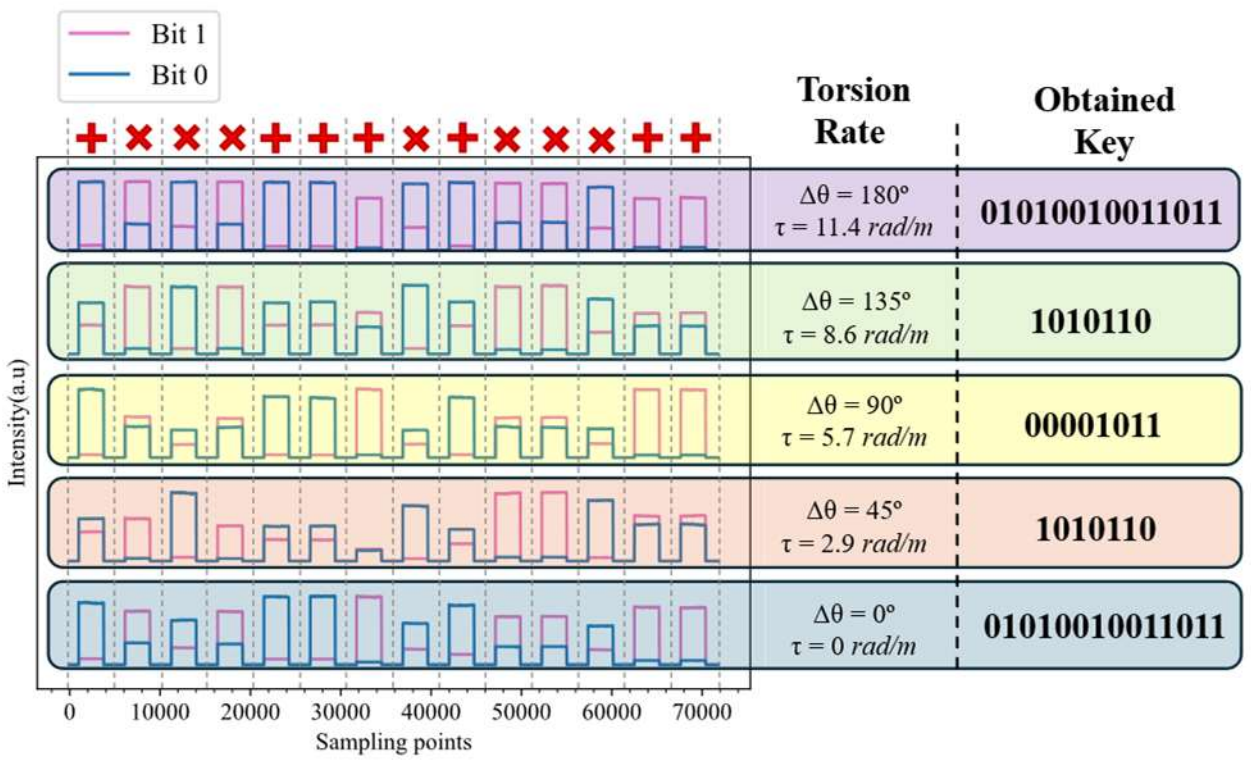

**Fig. 6.** Recovered key as a function of the torsion angle applied to the optical fiber.

To read the signal, the maximum value between the detectors was considered to select the corresponding bit. In order to evaluate the influence of torsion, the signal contrast, C, defined as $C = I_{max}/I_{min}$, was measured. When the contrast was greater than 2, indicating a difference of more than 50%, the selected bit was validated; otherwise, it was rejected and considered erased. Following this analysis, the *ER* was measured for each applied torsion value, as illustrated in Table III. When the number of erased bits exceeds a predefined threshold, the corresponding raw-key block should be discarded, and a new key-generation attempt should be performed after channel recalibration or polarization compensation.

TABLE III
*ER* Evaluation due to Torsion Influence

| Torsion Rate (*rad/m*) | *Recovered Key* | *ER* | *BER* |
|---|---|---|---|
| 0 | 01010010011011 | 0% | 0% |
| 2.9 | ×101×××0×110×× | 50% | 0% |
| 5.7 | 0×0×00××0×××11 | 50% | 0% |
| 8.6 | ×101×××0×110×× | 50% | 0% |
| 11.4 | 01010010011011 | 0% | 0% |

## V. CONCLUSION

This work experimentally investigated a classical BB84-inspired polarization-encoding scheme in an optical-fiber link under controlled temperature, strain, and torsion perturbations. For the tested bend-insensitive fiber, temperature variations from 24 °C to 35 °C and strain up to 2 mε caused negligible changes in the state of polarization and did not significantly affect bit recovery, indicating that these perturbations are not the dominant impairment within the tested ranges. In contrast, torsion strongly affected the transmitted state of polarization by modifying the ellipticity and orientation of the optical field, reducing detector intensity contrast, and causing basis-dependent degradation of the recovered key. From a communication perspective, torsion is therefore a critical impairment that must be monitored, compensated, or avoided in polarization-based BB84-inspired and quantum-enabled fiber links. From a sensing perspective, however, this torsion sensitivity can be exploited as a transduction mechanism, since the evolution of the state of polarization and the degradation of the recovered bit sequence provide measurable indicators of fiber twist. Thus, the proposed configuration can serve as a testbed for polarization-channel robustness and as a basis for future integrated communication-and-sensing systems, where channel perturbations are simultaneously detected and quantified.